# Chemical interaction and molecular growth of a highly dipolar merocyanine molecule on metal surfaces: A photoelectron spectroscopy study

Baris Öcal,[1] Philipp Weitkamp,[1] Klaus Meerholz,[1] and Selina Olthof [1*]

[1] Department Chemistry, University of Cologne, Greinstrasse 4-6, Cologne 50939, Germany

Email: solthof@uni-koeln.de

## Abstract

The growth and ordering of molecules on surfaces is an intriguing research topic as insights gained here can be of significant relevance for organic electronic devices. While often simple, rigid molecules are employed as model systems, we show results for a highly dipolar merocyanine which is studied on top of Au(100), Ag(100) and Cu(100) metal single crystals. Film thicknesses ranging from sub-monolayer to multilayer regimes are analyzed using UV (UPS) and X-ray photoelectron spectroscopy (XPS). For the monolayer regime, there is strong indication of face-on orientation, with both of the molecules' sulfur atoms bonding to the metal surfaces. Here, on Ag and Au(100) the sulfur atoms lose some or all of their intrinsic charges due to a charge transfer with the substrate, while on Cu(100) a strong metal-sulfur bond forms. The interaction between the substrate and the molecules can also be seen in the intensity and width of the highest occupied molecular orbital features in UPS. Upon multilayer deposition, a gradual lowering in ionization energy is observed, likely due to the formation of antiparallel dimers followed by an increased charge carrier delocalization due to the formation of an extended molecular aggregate for thicker layers. Interestingly, on Cu(100) the aggregated phase is already observed for much lower deposition, showing the importance of substrate-molecule interaction on the subsequent film growth. Therefore, this study offers a detailed understanding of the interface formation and electronic structure evolution for merocyanine films on different metal surfaces.

# 1. Introduction

Semiconducting organic molecules, suitable for optoelectronic applications, have been attracting significant interest in the past two decades. For these materials, the electronic and optical properties can be widely tuned by changing the molecule's structure or adding functional groups. When deposited as thin films, the properties, in particular with respect to charge transport, are strongly affected by the microstructure, which can vary from crystalline to fully amorphous.[1] Intriguingly, also the orientation of the molecule can matter for the electronic structure if molecules with a permanent dipole or quadrupole moment are employed. The orientation can lead to distinctly different values of ionization energy (IE)[2–4] which for example can affect the open circuit voltage of organic solar cells.[5] Also with regards to the absorption and emission properties, orientation and packing are of paramount importance.[6] In order to improve the performance of organic-semiconductor based devices, it is therefore desirable to be able to control the crystallinity and orientation of molecules in devices.

Achieving and investigating the ordered growth of molecules is therefore an intriguing research topic. In the past, particularly metal/organic interfaces have been investigated using surface sensitive techniques such as scanning tunneling microscopy (STM), low energy electron diffraction (LEED), near edge x-ray absorption fine-structure (NEXAFS), or photoelectron spectroscopy (PES).[7] Most of the molecules for which ordered growth has been studied have particularly rigid structures of high symmetry such as rubrene[8,9], PTCDA,[10,11] pentacene[12] or phthalocyanines.[13,14] Studying these types of molecules has helped scientists in the past to gain an understanding of the effects of molecule-molecule and molecule-substrate interaction on the ordered growth of films with a few monolayer thickness.[15,16] However, molecules currently employed in optoelectronic devices are typically more complex, having asymmetric structures and containing flexible side chains that are affecting the packing behavior. Furthermore, they typically incorporate various heteroatoms which can result in permanent dipole moments such as donor-acceptor (D-A) molecules or more complex structures, such as the non-fullerene absorbers having e.g. A-D-A structures.[17] These will affect the aggregation behavior and film growth. Only few groups have looked into the ordering on single crystal metal surfaces for such asymmetric or highly dipolar molecules.[18,19]

Merocyanine molecules contain donor and acceptor groups that are connected with a π-conjugated bridge, which results in a dipole moment that can be tuned by the choice of D and A sub-units.[20] An example of the merocyanine molecule (2-[5-(5-dibutyl-amino-thiophene-2-yl-methylene)-4-tert-butyl-5H-thiazol-2-ylidene]-malononitrile (HB238), relevant for the present work, is shown in Figure 1. This class of molecules is known for their high dipolar moment, high extinction coefficients and flexibility of their chemical structure, which makes them intriguing candidates for organic solar cell applications.[21] These molecules tend to form aggregates in thin films, yielding good charge transport properties, which makes them also of interest for organic transistors.[22] Overall their unique intermolecular interactions makes them attractive candidates to studying their optical [23,24] and electronic [25,26] properties.

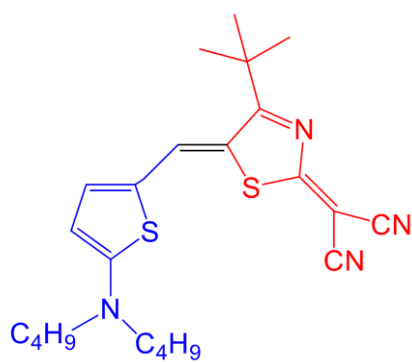

Figure 1: Molecular structure of the merocyanine HB238. The donor (D) sub-unit is indicated by blue color and the acceptor (A) is shown in red.

Already around 30 years ago, Seki et al. performed UV photoelectron spectroscopy (UPS) measurements to investigate merocyanines on silver halides,[27] which were utilized as sensitizers in silver halide photography. They found large interface dipoles (~1 eV) that are formed within the first ~ 1 nm thickness. This suggests that these merocyanines aligned with negative charge on the substrate and positive towards vacuum. Additional NEXAFS studies of various merocyanines showed that the dye molecules containing a thioketone C=S sites adsorbed strongly with this sulfur to the metal halide surface resulting in an edge on orientation[28] which led to the formation of a dipole layer. In contrast, sulfur bound in a thiole group was little or not participating in bonding. For larger thicknesses a more random orientation occurred which canceled out the effect of the dipole moment.

Furthermore, Koch et. al investigated a range of different merocyanines on a Au(111) single crystal [26] as well as metal oxides[29] by UPS. On Au(111) no correlation between the molecule's dipole moment and a changes in work function (Wf) was found, which suggested flat lying molecules in the first monolayer. Furthermore, they found a significant change in the position and shape of the highest molecular orbital (HOMO) position between first and second layer, which indicated that molecules dimerize for thicknesses larger than a monolayer. In contrast, on the metal oxide only weak coordination bonds formed between the electron withdrawing carbonyl or cyano groups with the surface metal atoms, leading to a vertical arrangement and a strong dipolar interface layer.[30]

A more recent study using STM and LEED showed that a sub-monolayer of the merocyanine HB238 on Ag(100) shows an commensurate chiral structure with face on orientation, which confirms that indeed the molecule shown in Figure 1 can form ordered thin films.[31]

Based on these earlier works it is already evident that the arrangement of merocyanines on a surface is highly substrate dependent. In this study, we want to investigate the electronic structure of the same highly dipolar merocyanine HB238 (dipole moment of 13.1D [32]) on three different metal substrates to further investigate the substrate-molecule interaction in the monolayer region and its impact on the subsequent film growth for thicker layers. Here, Au(100), Ag(100), and Cu(100) are chosen, due to their different chemical affinities. Using UPS as well as x-ray photoelectron spectroscopy (XPS) we find that the chemical interaction strength significantly affects the monolayer electronic structure and that both sulfur atoms (see Figure 1) are participating in the chemisorption, indicating flat lying molecules in the first monolayer. Subsequent multilayer deposition leads to significant changes in IE, due to changes in orientation, dimerization and aggregate formation as the molecule decouple from the surface. The molecule shows strong hybridization only on Cu(100) followed by a different growth mode in comparison to Ag and Au(100), which suggests that substrate molecule interactions at the monolayer regime can significantly alter the molecular growth.

## 2. Experimental Section

The substrate preparation and photoelectron spectroscopy measurements have been performed under ultra-high vacuum conditions ($< 10^{-8}$ mbar). The metal single crystals Ag(100), Au(100), and Cu(100) (MaTeck GmBh) were cleaned by multiple sputter and annealing cycles. Each Ar sputtering has been performed at 2 kV for 15 min and afterwards the samples were annealed at 550°C for 2 hours with radiative heating. The processes is repeated until a clean XPS spectrum is obtained, that only shows negligible traces of carbon residuals. After substrate preparation, HB238 is evaporated using a Knudsen cell (Creaphys) at a rate of approximately 0.01 Å/s, as determined by a crystal microbalance using a density of 1.6 g/cm$^3$. The sample is kept at room temperature during evaporation. Sample transfer through the multichamber setup into the measurement chamber is done without breaking the vacuum.

For the UPS measurements, a monochromated helium excitation lamp was used (VUV 5k, Scienta Omicron), tuned to the HeI$_\alpha$ line at an energy of 21.22 eV. Note that the monochromator was slightly misaligned to reduce the intensity of the incoming UV light by a factor of approximately 4 to minimize degradation of the molecules, which is known to be an issue for merocyanine molecules.[26] Nonetheless, after 42 Å thickness, a slight shift in spectra can be seen under extended UV-light exposure and shown in Supporting Materials Figure S1, therefore no measurements beyond that layer thickness are included in this study. XPS was performed on the same samples as UPS using a non-monochromatic dual anode x-ray source (VG) to investigate changes in the binding energy of the N1s, C1s and S2p core levels. For the interfaces on the Ag(100) and Au(100) the MgK$_\alpha$ line at 1254.64 eV was used, while for the Cu(100) sample this was changed to AlK$_\alpha$ with an energy of 1486.61 eV as otherwise the Auger lines of copper would have coincided with the N1s core level region. A hemispherical analyzer (Specs, Phoibos 100) at pass energies of 30 eV for XPS core level scans and 2 eV for UPS scans has been used to detect the emitted photoelectrons. It was confirmed that the XPS measurements do not damage the molecules, by checking with UPS before and after the extended (~ 6 h) XPS measurements.

The XPS fits have been made with the fitting program *XPSPeak41*. FWHM and Lorentian to Gaussian ratios have been determined for the S2p and N1s signals and were kept constant for all thicknesses. A Shirley function was used to subtract the background. The distance between spin orbit split S2p$_{1/2}$ and S2p$_{3/2}$ was set to 1.23 eV and the area ratio of the donor and acceptor S2p peaks were fixed to 1:1. For the N1s peaks the area ratio was set to 1:1:2 due to molecules stoichiometry. Depending on the exact choice of fitting parameters, the peak positions can be shifted by around 0.05 - 0.1 eV, which can be therefore considered as the error bar of the analysis.

XRD measurements of 30 nm HB238 on the different single crystals have been performed using an Empyrean (Malvern Panalytical) system equipped with a non-monochromatic Cu K$_\alpha$ anode. A 1/32° divergence slit was used for all measurements. The samples were transferred and measured in air, though the exposure time was limited to < 1h to minimize possible degradation.

## 3. Results and discussion

### 3.1 Investigations by UPS

To investigate the film properties of the merocyanine HB238 on the three single crystals Au(100), Ag(100) and Cu(100), UPS measurements were conducted. Here various film thickness were studied, ranging from below a monolayer coverage (approximately 2-3 Å) up to the multilayer regime (42 Å). The full UPS spectra are included in the Supplementary Materials

Figure S2, while the extracted values of the work function and ionization energy are summarized in Figure 2a) and b), respectively. This data provides insights on how the electronic structure is changing with layer thickness, and therefore how substrate-molecule and molecule-molecule interaction change depending on the choice of substrate as well as layer thickness.

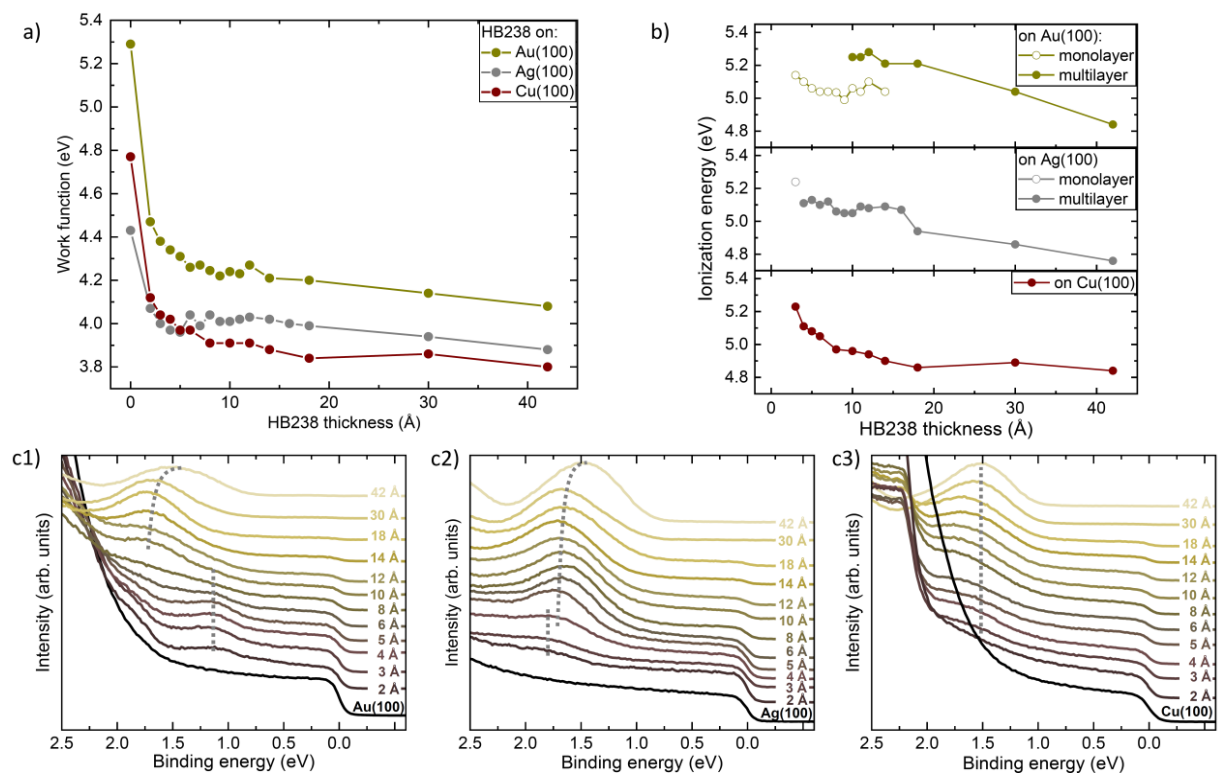

Figure 2: Data obtained from the thickness dependent UPS measurements on the three substrates Au(100), Ag(100) and Cu(100), the full spectra are shown in the Supplementary Materials, Figure S2. In a) the extracted values for the work function are shown, while b) summarizes the changes in ionization energy. (c1) to (c3) show a selection of data sets for the HOMO regions on the three crystals. The spectra of the different layer thicknesses were shifted vertically for clarity. The dashed lines act as a guide for the eye to indicate changes in HOMO peak position.

The first data points in Figure 2a) at 0 Å coverage represent the Wf's of the bare substrates, where we find 5.29, 4.43 and 4.77 eV for Au(100), Ag(100) and Cu(100), respectively. Upon 2 and 3 Å of deposition of the molecule, the Wf values rapidly decrease on all three metals by $\Delta Wf(Au) = 0.91$ eV, $\Delta Wf(Ag) = 0.43$ eV and $\Delta Wf(Cu) = 0.69$ eV. This behavior is well known for metals and is usually attributed to the so-called push-back effect in which the electrons spilling out of the metal surface are pushed back into the metal bulk in presence of adsorbed material which affects the Wf.[33,34] The change of approximately 1 eV on Au(100) fits previously reported values for the push-back effect on this metal, where the effect was attributed to physisorbed molecules.[35]

After 3 Å coverage, all three samples show a change in the slope of the Wf decrease, therefore it can be assumed that at this thickness a monolayer of molecules is fully covering the metals. As this change in slope happens after the same coverage on all three surfaces, and it is known from previous STM studies of HB238 on Ag(100) that the first monolayer of molecules is lying face-on on the surface,[31] this indicates that all three metals exhibit flat-lying molecules in the first monolayer. This is also in agreement with the previous study by Koch et al., where such an orientation was suggest on Au(111).[26] Further insights into this interface formation will be discussed below in the context of the XPS measurements.

For Au(100) and Ag(100), between 3 Å and approximately 5 - 7 Å the work functions decrease by approximately 0.1 eV, before remaining rather stable within the experimental error until 18 Å. However, between 18 and 42 Å there is another decrease in Wf for these two substrates by 0.3 – 0.4 eV. In contrast, for Cu(100) we observe all of the gradual work function decrease by 0.4 eV between 3 and 18 Å, after which the value remains mostly constant until 42 Å. Notably, for the 42 Å thick layers on all three substrates the final Wf values are similar. The underlying effects leading to these variations in work function change beyond the monolayer regime are not straight forward to comprehend, but likely related to changes in molecular orientation and packing[36], as discussed next.

More insights can be gained from the analysis of the changes in IE, since this value is affected e.g. by the orientation of the molecules, the polarizability of the surrounding, and the hybridization of the orbitals. The change in IE value as a function of layer thickness is plotted in Figure 2b). Note that below a monolayer coverage this value is not defined, as the work function will be affected by the presence of the bare surface,[37] so the first data point is given for 3 Å coverage. In addition to the IE values, Figure 2c1), c2) and c3) also provide the detailed views of the UPS spectra in the low binding energy regions, where the changes in the HOMO shape can be observed.

First, the observations for Au(100) will be discussed. Here, a HOMO peak with an onset position at 0.83 eV can be seen upon the deposition of a monolayer. The peak is rather narrow with a full width half max (FWHM) of 0.43 eV. Upon increasing the thickness, a second much broader peak emerges, which becomes dominant at around to 10 Å, where we observe an onset position at 1.15 eV and a FWHM of 0.65 eV. In the previously mentioned work by Koch et. al on Au(111) this broader HOMO peak was attributed to molecules in an antiparallel packing (dimers) and edge-on orientation.[26] Though the initial narrow HOMO feature becomes suppressed with increasing coverage, this monolayer peak can still be seen up to 12 Å layer thickness and remains at the same position. This indicates that on Au(100) the lying down molecules of the first monolayer are not affected by the subsequent multilayer growth of the dimers. Due to the presence of the two distinct HOMO features, the IE plot in Figure 2b) contains two IE values, one for the monolayer (open symbols) which remains at ~ 5 eV, and one for the multilayer (filled symbols). Between 10 Å and 18 Å, this multilayer HOMO signal is becoming more intense whereby the extracted values of IE ≈ 5.2 eV as well as the FWHM and HOMO onset do not change. Therefore, from the second layer on the molecules are growing in the same orientation and packing motive, which is the reason why the work function is also not changing in this regime. After 18 Å, the HOMO gradually shifts to lower binding energy by ~ 300 meV, the FWHM increases from 0.65 to 0.72 eV and the ionization energy decreases from 5.34 to 4.84 eV. This is the same regime in which the work function is also starting to decrease again. The lowering of the ionization energy could be the result of a change in molecular orientation, or of the layer becoming more amorphous. [36,38] Alternatively, an extended crystalline layer could exhibit increased coupling[39,40] which in turn affects the delocalization of charge carriers and thereby the polarization screening during the UPS measurement. This would also lead to a lowering of the IE, an effect known as final state screening.[41,42]

To test for the presence or absence of a distinct orientation with increasing layer thickness, we performed x-ray diffraction (XRD) measurements of a 30 nm HB238 layer; the data is included in the Supplementary Material Figure S3a. Here, in addition to Kiessig oscillations within the thin organic layer also a single pronounced reflex is found at 2Θ ≈ 5°. This shows that the evaporated layer on Au(100) is highly ordered with only one crystalline phase presence, which corresponds to a tip-on orientation for aggregated antiparallel aligned HB238 molecules, as previously reported.[43,44] Results obtained by optical spectroscopy of such layers have indicated that here the excitons are delocalization over multiple molecules. [45,46] We therefore suggest that the lowering in Wf, and more importantly the decrease in IE observed on Au(100), comes from the increased molecule-molecule interaction in the extended ordered layer which broadens the HOMO, delocalizes the charge carriers and increases the polarization screening.

On Ag(100), a broader HOMO feature (FWHM of 0.66 eV) is found already for the monolayer regime, which is located at a higher binding energy of 1.33 eV compared to the first layer on Au. The higher binding energy and peak width indicate stronger hybridization of the molecular orbitals with the substrate, which will be further discussed based on the XPS results presented below. When increasing the thickness beyond the monolayer regime to 5 Å, a second broader HOMO feature arises at 1.06 eV with a FWHM of 0.73 eV; the extracted value of IE changes from 5.23 for the monolayer to 5.11 eV for the multilayer. The monolayer peak cannot be distinguished after 3 Å coverage, as the multilayer peak overlaps with it. Based on the IE value as well as the onset of the HOMO, the multilayer closely resembles the second phase on Au(100). The changes in energy levels for increasing layer thickness are also reminiscent of the measurement on Au(100). While initially being constant, from 18 Å on again an increase in the HOMO's FWHM, a shift of the HOMO towards lower binding energy and a decrease in IE to 4.76 eV is observed. It is therefore likely, that the growth and orientation on Au and Ag(100) is similar, which is supported by the corresponding XRD results on the 30 nm thick molecule layer on Ag(100). Here, as seen in Figure S3b of the Supplementary Materials, similar Kiessig oscillations and a reflex at $2\Theta \approx 5°$ is found.

In contrast to the other two substrates, on Cu(100) a HOMO feature is not detectable for the monolayer regime. It is possible that the HOMO peak of the molecule is coinciding with (and therefore masked by) the pronounced valence band states of the Cu(100) substrate, but for this it would have to be located at much higher binding energy compared to the other substrates. More likely, the interaction between the Cu surface and the molecules is so strong that the π-electron system, and thereby the energy levels, are disturbed. This is indeed what the XPS measurements shown below are suggesting. Starting from 4 Å on, a broad HOMO peak at 1.06 eV with a FWHM of 0.61 eV can be observed. Between 4 and 18 Å the ionization energy gradually lowers from ~ 5.0 eV to ~ 4.8 eV. After 18 Å coverage, neither the position nor the FWHM of the HOMO change any more, and correspond to values found for the thickest 42 Å layer on Au and Ag(100). This indicates that on Cu(100) the molecules aggregates more quickly that on the noble metals. The XRD measurements in the Supplementary Material Figure S3c show the same orientation and crystal phase, but notably a lack of Kiessing fringes. This indicates a rougher film[47] and therefore likely on Cu(100) island growth takes place after the initial monolayer coverage, while on Au and Ag(100) the molecular films exhibit a layer-by-layer like growth.

## 3.2 Investigations by XPS

In order to better understand the variations seen in Wf and IE for the thickness dependent UPS data discussed above, XPS measurements have been conducted to probe changes in core level binding energies of the HB238-specific atoms on the three different substrates. These measurements can give insights into changes in donor-acceptor strength (i.e. dipole moment)[31] and indicate which atoms bind to the substrate.

The XPS core level signals of the merocyanine molecule under investigation here can be expected to be rather complex, with the heteroatoms C, N and S being present in different chemical bonds; therefore multiple peaks will contribute to the overall XPS core level signals in all cases. Moreover, due to the strong dipole moment, even similar bonds (e.g. C=C) will appear at distinctly different binding energies depending on whether the atoms are included on the donor or acceptor side of the molecule. For the carbon peak, the resulting fit would likely include more than ten distinct bonds and is too complicated to yield meaningful results; it is therefore omitted in this discussion.

The fitting of the sulfur peaks is more straightforward. One sulfur is located on the donor part of the molecule in a thiophene ring and the other one in the acceptor subunit in a thiazole group (see Figure 1). Though the chemical environment can be considered to be similar in both cases, due to the inherent dipole moment of HB238 these heteroatoms are differently charged and should appear at different binding energies in an XPS measurement. The difference in binding energy can therefore be an indication of the dipole moment of the molecule. In addition sulfur is known to show strong chemical affinity to various metals,[48] therefore analyzing the changes in S2p position can be very insightful when studying the interface formation.

The fitted S2p spectra of the interfaces between HB238 and the three single crystal surfaces are shown in Figure 3. Here, the fitted contribution by the S2p doublet originating from the donor subunit is shown in blue color, while the acceptor is given in red. For the thickest films (42 Å) we find a very similar fit on all three substrates, where in addition to these two sulfurs doublets, also a contribution by a shake-up peak is seen at higher binding energy, marked by a diamond symbol, originating from π → π* transitions. However, for thinner layers there are distinct differences fond for the three substrates, which will be discussed next.

For all three substrates, the fits in Figure 3 reveal that the distance between the two sulfur contributions decreases close to the interface, i.e. for low coverages. The values for the donor and acceptor $S2p_{3/2}$ peak positions are summarized in Table S1 in the Supplementary Materials. The XPS spectra on Au(100) are shown in Figure 3a1) and the peak positions as a function of layer thickness are plotted in a2). The acceptor and donor peaks of the 2 Å measurement are located at 164 and 164.32 eV, respectively. For increasing film thickness the sulfur donor atoms become more positively charged (binding energy increases from 164.32 to 164.51 eV for the 11 Å film) while the acceptor sulfur is becoming more negatively charged (binding energy decreases from 164 to 163.85 eV). Due to this opposing shifting, the separation of the two sulfur peaks increases from $\Delta E_{bind}$ = 0.32 eV for the 2 Å layer to $\Delta E_{bind}$ = 0.64 eV for 11 Å thickness. It is therefore likely, that the molecules have a face-on orientation for the (sub-)monolayer regime as suggested already above, whereby the positive donor and negative acceptor heteroatoms charges are reduced (screened) by charge transfer. From the second layer on, the sulfur atoms are not affected by the metal anymore, the molecules are decoupled from the surface. A gradual shifting is observed in the extracted binding energies, instead of an abrupt transition, due to the fact that the measurement, and therefore the fit, is an average of the mono and multilayer peaks due to the probing depth of XPS of several nm. With further increasing thickness from 11 to 42 Å, the S peak separation stays approximately the same, but both S peaks shift to higher binding energy by ~ 150 meV. In the UPS related data shown in Figure 2a), we observed a similar downward shift of the vacuum level, i.e. a reduction in Wf, by the same amount; therefore, the change in Wf of the layer is the origin of the S2p core level shift observed here.

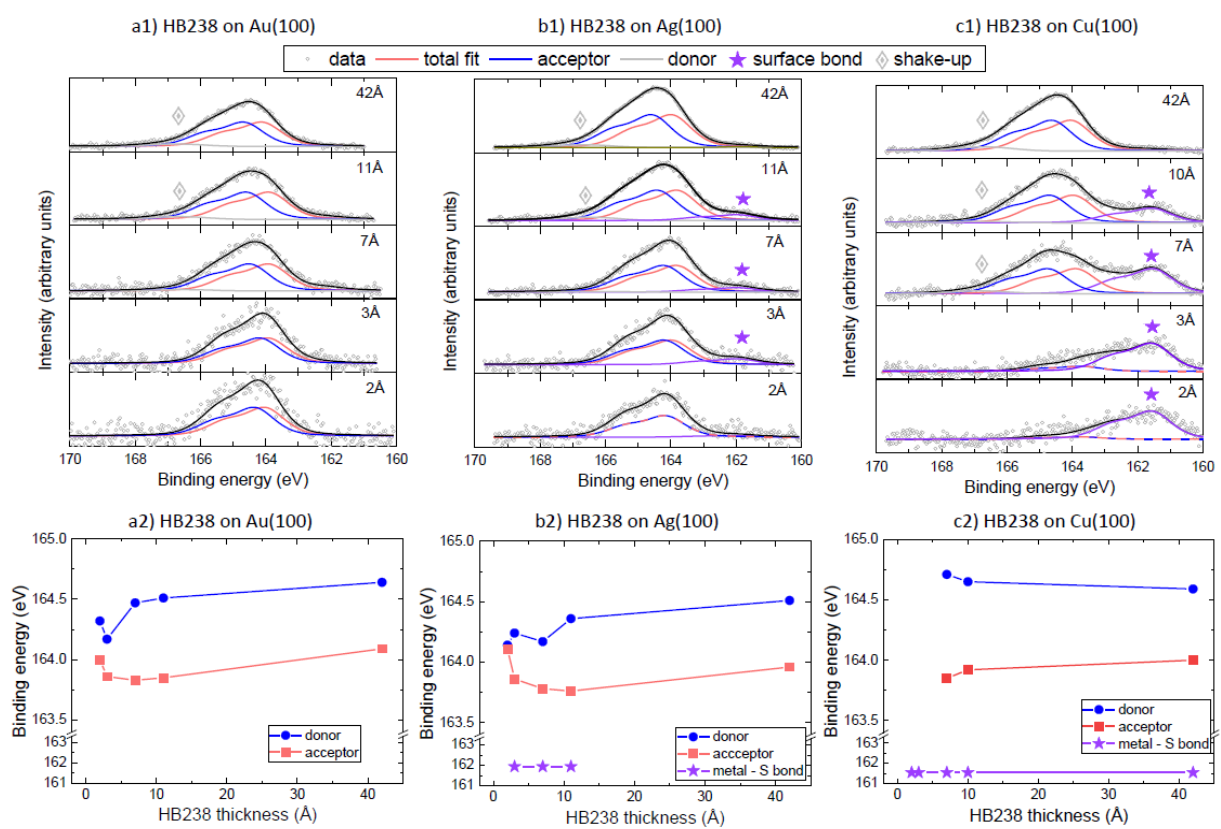

Figure 3: XPS measurements of S2p core level signals of HB238 at different thicknesses on a1) Au(100), b1) Ag(100) and c1) Cu(100). The red and blue fits correspond to the contributions from the donor and acceptor sulfur, respectively. Additional features from a substrate bond (indicated by star) and shake-up peaks (diamond) are marked. The extracted positions of the S2p$_{3/2}$ peaks as a function of HB238 thickness are shown in a2), b2) and c2).

In the case of the Ag(100) interface, the 2 Å of HB238 data shows no difference between donor and acceptor S2p core levels, both are located at 164.1 eV, which is similar to the average location of the peaks observed on Au(100). Again, the molecule must have a face-on orientation with both sulfur atoms close to the surface, but this time with a stronger metal – molecule interaction compared to the case of Au(100), which neutralizes the charges on the sulfur atoms due to charge transfer. The result explains also the broadening of the HOMO at the monolayer regime for Ag(100) which is attributed to the stronger chemisorption. The measurements for 3 to 11 Å show a weak additional S2p doublet at lower binding energy of 161.94 eV; the feature is marked by a star in Figure 2b). As this peak coincides with reports on chemisorbed sulfur, we suspect that a small fraction of the molecules are adsorbed at defect sites or step edges of the substrate where they encounter a different binding environment. The changes in binding energy are summarized in Figure 3 b2) and the behavior for increasing film thickness resembles the one on Au(100). The donor sulfur peaks become more positively charged (binding energies increase from 164.15 to 164.36 eV) while acceptor sulfur becomes more negatively charged (binding energy decreases from 164.1 to 163.76 eV); after 11 Å deposition, again a separation of approximately $\Delta E_{bind} \approx 0.6$ eV is reached. The difference in the peak shifts suggests that the molecule becomes slightly positively charged on the surface. Also here, between 11 Å and 42 Å a work function related shift of the core levels by ~0.15 eV is observed to higher binding energies, similar to the case on Au(100). Comparing the 11 and 42 Å thick films on both Au and Ag, there is a difference of 0.1 eV in S2p$_{3/2}$ binding energy, which is in good agreement with the difference observed in work function by UPS.

On Cu(100), the fit for the lower layer thicknesses looks distinctly different. For the 2 Å layer a pronounced signal at lower binding energy is observed, indicated by a star in Figure 2 c1). Its binding energy at 161.6 eV can be assigned to strong Cu-S bonds.[49] Obviously, the chemical affinity between Cu and S is much higher than for the two noble metals and S. Clearly, also on this surface the molecules must be lying flat since in an edge-on orientation is not possible for both S atoms to bind to the Cu atoms on the surface. In addition to this surface bond, a weak feature is seen around 164.2 eV, therefore similar in position to the one observed on silver and gold. The origin of these more weakly bound molecules, which are also seen for the 3 Å measurement, is unclear; possibly these are molecules adsorbed on contaminated / passivated areas of the surface. As the signal is low, it will not further be discussed for the sub-monolayer regime. The XPS measurements reveal that on Cu(100), the molecule in the first monolayer is much more strongly bonded which is the reason why in the UPS measurement in Figure 2 c3) no HOMO signal was detectable. Once the multilayer forms and the molecules decouple form the surface, the sulfur surface bond at lower binding energy gets rapidly suppressed and the expected signals from the donor and acceptor sulfur emerge at higher binding energy. For the 7 Å coverage there already is a distance of $\Delta E_{bind} \approx 0.8$ eV between donor and acceptor S2p peaks before reducing to the expected value of $\Delta E_{bind} \approx 0.6$ eV for the 11 and 42 Å data set. As there is no reason for the 7 Å film to have a larger S peak separation, this is likely an inaccuracy of the fitting procedure. For the 11 Å and 42 Å data no changes in S acceptor and donor peak locations are seen, they remain at 164 and 164.6 eV, respectively; for these layer thicknesses, no change in Wf in the UPS data was observed. Notably, these positions are in the same range as were found on Ag(100) and Au(100) and therefore on all three samples there is an identical distance between these donor and acceptor peaks. Based on the XPS measurements, the multilayer behavior on Cu(100) is similar to on Ag(100) and Au(100) even though the monolayer is more strongly chemisorbed.

So far, the changes in sulfur peak position have been discussed, which are strongly affected by the chemisorption strength and charge transfer at the metal interface, and are therefore not necessarily representative of changes observed by the molecule as a whole. Next, we analyze the N1s signal of all data sets; the XPS N1s core level spectra as well as the extracted peak positions are summarized in Figure 4 and the binding energy values are summarized in Table S2 in the Supplementary Materials. Looking at the molecule's structure in Figure 1, we expect to observe N1s signals at three distinct binding energies in a ratio 2:1:1. These are the two nitrogen atoms in the cyano group (C≡N) of the acceptor which are labeled as N(1) in the following, the nitrogen in the acceptor's thiophene ring labeled as N(2) and the nitrogen connected to the alkyl chains in the donor labeled as N(3). It should be noted that fitting of the N1s region in the case of the Ag substrate is challenging for the thinnest layers, since a plasmon loss feature, associated with the Ag3d core level, extends into this region. For the data shown in Figure 4 b1) this plasmon background has been subtracted; the original data, including the process for background subtraction, are shown in the Supplementary Materials Figure S4. Also for the thick layers on all three substrates there is another uncertainty in the fit, since only one shake-up peak is considered, though all three peaks can be accompanied by such a feature. But since the intensity and position of the individual shake peaks is not known, such a fit with 6 components leads to less reproducible and meaningful results.

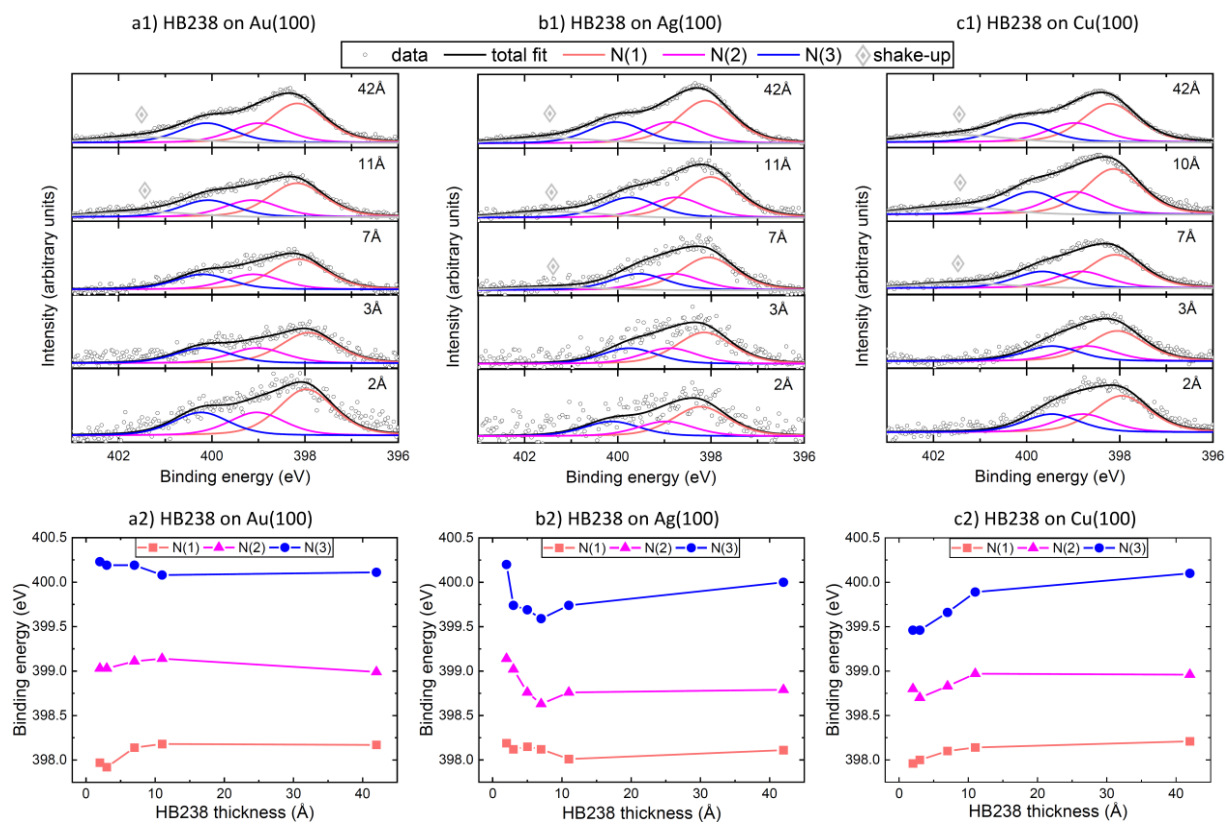

Figure 4: XPS measurements of N1s core level peaks of HB238 at different thicknesses on a1) Au(100), b1) Ag(100) and c1) Cu(100) with the corresponding fit of the various nitrogen species in the molecule. The extracted positions of the different N1s species as a function of layer thickness are plotted in a2) for Au(100), in b2) for Ag(100) and in c2) for Cu(100).

On Au(100), the nitrogen peak positions for the monolayer regime are 397.97 eV and 399.03 eV for the acceptor based nitrogen N(1) and N(2), as well as 400.23 eV for the donor-based N(3), as plotted in Figure 4 a2). The fact that the binding energy of N(2) and N(3) are different by 0.9 eV indicates that also for the monolayer of flat lying molecules some of the merocyanine's dipole moment remains, since otherwise these two nitrogen with their similar chemical environment (C-N bonds) should have similar peak positions. With increasing thickness from 2 to 11 Å, the N(3) binding energy slightly decreases, while this value slightly increases for N(1) and N(2). Therefore, the difference in binding energy for the nitrogen atoms in the donor and acceptor part for molecule lowers as the molecules decouple from the surface from the second monolayer on. Notably, this trend is opposite to what was observed for the S2p peaks, where peak positions were dominated by the surface bond and direct charge transfer. The decrease in separation from approximately $\Delta E_{bind} \approx 1.2$ eV at 2 Å to 0.9 eV at 11 Å of the two chemically similarly bound nitrogen N(2) and N(3) indicates a lowering of the molecule's effective dipole moment with increasing layer thickness, which is likely due to the formation of the dimers which shields the overall dipole. The results are consistent with our UPS evaluation, where a second broad HOMO peak is observed at higher binding energies at these thicknesses, which we already have associated with these dimers. Small changes in peak position (< 0.2eV) are also seen between 18 and 42 A. Whether these minor changes are representative, or a result of the fitting uncertainties is not clear.

On Ag(100), the N peak positions for the monolayer are similar to the measurement on Au(100) considering the error margin of the fitting, even though here the S2p peaks showed considerable difference in binding energy due to the stronger bond to the Ag metal surface. As shown in Figure 4 b2), when the thickness is increased from 2 to 11 Å, all N1s signals shift to lower binding energy. This suggest that at the monolayer regime the molecule is slightly positively charged on Ag(100); a similar observation was made for the S2p peaks where the donor S2p became more positively charged on the surface compared to the acceptor which was more negatively charged. With increasing film deposition, the shift in binding energy is stronger for the acceptor nitrogen N(2) and N(3) compared to the donor N(1), therefore also here the dipole is screened through the formation of dimers. When the thickness is further increased to 42 Å, all peaks show a slight increase in binding energy and the final positions are similar to the results obtained for Au(100).

For the thinnest layer of 2 Å on Cu(100), the N(3) donor is located at 399.46 eV, while the two acceptor nitrogen are at 398.8 and 397.76 eV for N(2) and N(1), respectively. In this case, all N1s binding energies are significantly lower in comparison to the noble metals, due to strong charge transfer from the substrate to the whole molecule as discussed for the S2p peaks; this charges the molecule negatively. For thicknesses up to 11 Å, the peaks shift to higher binding energy; here the N(3) binding energy increases around 0.5 eV, while N(2) and N(1) increase around 0.25 eV. Notably, the separation in binding energy of N(2) and N(3) is smaller for the monolayer ($\Delta E_{bind} \approx 0.7$ eV) compared to the 11 Å film ($\Delta E_{bind} \approx 0.9$ eV, similar to the noble metals). Therefore, the strong charge transfer on the Cu(100) surface shields the molecules' dipole moment more effectively in the monolayer compared to the dimerization in the multilayer film. Again the final peak positions are similar to the ones observed on Au and Ag(100) at 42 Å thickness.

## 4. Conclusion

The effect of chemisorption strength on the layer growth of a highly dipolar merocyanine molecule has been investigated by photoelectron spectroscopy using three single crystal metal surfaces with different chemical affinity of Au(100) < Ag(100) < Cu(100). The XPS results show that in the first monolayer the molecules bind via both sulfur atoms on all three substrates and are therefore likely lying flat on the surface. Hereby, the bonding strength of the molecule on the noble metals is overall rather weak, but a slight effect on the width of the HOMO feature measured in UPS can be seen, which is more narrow on Au(100) compared to the more strongly hybridizing Ag(100). In contrast, on Cu(100) a strong sulfur-metal bond is found, leading to a pronounced shift in the S2p core level and a strong hybridization of the π-electron system for the monolayer whereby the HOMO feature is fully suppressed. From the second layer on, on all substrates a broader HOMO feature emerges, which is associated with the formation of antiparallel dimers. On Au(100) and Ag(100) this initial dimer layer has an IE of ~5.2 eV, while for thicknesses beyond ~18 Å this value gradually decreases to 4.8 eV due to the increased molecule-molecule interaction and polarization screening in the extended ordered film. On Cu(100) this low IE phase is already reached at much lower coverage, which is likely due to island growth which forms larger aggregates already at an earlier stage. Our detailed investigation shows how UPS and XPS can give insights in the substrate-molecule as well as molecule-molecule interaction and helps us to better understand the growth of more complex organic semiconductors on such ordered surfaces.

## CRediT authorship contribution statement

**Öcal Baris:** Writing – original draft, Conceptualization, Investigation, Formal analysis, Data curation  **Weitkamp Philipp:** Writing – original draft, Data curation, Formal analysis **Meerholz Klaus:** Project administration, Funding acquisition, Resources, Supervision **Olthof Selina:** Funding acquisition, Writing – review & editing, Visualization, Data curation, Supervision, Conceptualization.

## Declaration of competing interest

The authors declare that they have no known competing financial interests or personal relationships that could have appeared to influence the work reported in this paper.

## Data availability

Data will be made available on request

## Acknowledgments

The authors thank Dirk Hertel for assisting with the XRD measurements as well as Moritz Sokolowski for fruitful discussions. This work was financially supported by the DFG funded Research Training Group RTG259 "Template-Designed Organic Electronics (TIDE)" as well as the excellence initiative of the University of Cologne, key profile area "Quantum Matter and Materials" (QM2).

# References

[1] H. Bässler, A. Köhler, Charge transport in organic semiconductors, Top. Curr. Chem. 312 (2012) 1–65. https://doi.org/10.1007/128_2011_218.

[2] S. Duhm, G. Heimel, I. Salzmann, H. Glowatzki, R.L. Johnson, A. Vollmer, J.P. Rabe, N. Koch, Orientation-dependent ionization energies and interface dipoles in ordered molecular assemblies, Nat. Mater. 7 (2008) 326–332. https://doi.org/10.1038/nmat2119.

[3] N. Koch, I. Salzmann, R.L. Johnson, J. Pflaum, R. Friedlein, J.P. Rabe, Molecular orientation dependent energy levels at interfaces with pentacene and pentacenequinone, Org. Electron. 7 (2006) 537–545. https://doi.org/10.1016/j.orgel.2006.07.010.

[4] G. Koller, S. Berkebile, J. Ivanco, F.P. Netzer, M.G. Ramsey, Device relevant organic films and interfaces: A surface science approach, Surf. Sci. 601 (2007) 5683–5689. https://doi.org/10.1016/j.susc.2007.06.070.

[5] U. Hörmann, C. Lorch, A. Hinderhofer, A. Gerlach, M. Gruber, J. Kraus, B. Sykora, S. Grob, T. Linderl, A. Wilke, A. Opitz, R. Hansson, A.S. Anselmo, Y. Ozawa, Y. Nakayama, H. Ishii, N. Koch, E. Moons, F. Schreiber, W. Brütting, Voc from a morphology point of view: The influence of molecular orientation on the open circuit voltage of organic planar heterojunction solar cells, J. Phys. Chem. C 118 (2014) 26462–26470. https://doi.org/10.1021/jp506180k.

[6] Q. Li, Z. Li, Molecular Packing: Another Key Point for the Performance of Organic and Polymeric Optoelectronic Materials, Acc. Chem. Res. 53 (2020) 962–973. https://doi.org/10.1021/acs.accounts.0c00060.

[7] R.J. Maurer, V.G. Ruiz, J. Camarillo-Cisneros, W. Liu, N. Ferri, K. Reuter, A. Tkatchenko, Adsorption structures and energetics of molecules on metal surfaces: Bridging experiment and theory, Prog. Surf. Sci. 91 (2016) 72–100. https://doi.org/10.1016/j.progsurf.2016.05.001.

[8] H. Ding, Y. Gao, Electronic structure at rubrene metal interfaces, Appl. Phys. A Mater. Sci. Process. 95 (2009) 89–94. https://doi.org/10.1007/s00339-008-5038-5.

[9] J.A. Miwa, F. Cicoira, J. Lipton-Duffin, D.F. Perepichka, C. Santato, F. Rosei, Self-assembly of rubrene on Cu(111), Nanotechnology 19 (2008). https://doi.org/10.1088/0957-4484/19/42/424021.

[10] S. Duhm, A. Gerlach, I. Salzmann, B. Bröker, R.L. Johnson, F. Schreiber, N. Koch, PTCDA on Au(1 1 1), Ag(1 1 1) and Cu(1 1 1): Correlation of interface charge transfer to bonding distance, Org. Electron. 9 (2008) 111–118. https://doi.org/10.1016/j.orgel.2007.10.004.

[11] S. Mannsfeld, M. Toerker, T. Schmitz-Hübsch, F. Sellam, T. Fritz, K. Leo, Combined LEED and STM study of PTCDA growth on reconstructed Au(1 1 1) and Au(1 0 0) single crystals, Org. Electron. 2 (2001) 121–134. https://doi.org/10.1016/S1566-1199(01)00018-0.

[12] P.G. Schroeder, C.B. France, J.B. Park, B.A. Parkinson, Energy level alignment and two-dimensional structure of pentacene on Au(111) surfaces, J. Appl. Phys. 91 (2002) 3010–3014. https://doi.org/10.1063/1.1445286.

[13] Z.H. Cheng, L. Gao, Z.T. Deng, Q. Liu, N. Jiang, X. Lin, X.B. He, S.X. Du, H.J. Gao, Epitaxial growth of iron phthalocyanine at the initial stage on Au(111) surface, J. Phys. Chem. C 111 (2007) 2656–2660. https://doi.org/10.1021/jp0660738.

[14] S. Ahmadi, M.N. Shariati, S. Yu, M. Göthelid, Molecular layers of ZnPc and FePc on Au(111) surface: Charge transfer and chemical interaction, J. Chem. Phys. 137 (2012). https://doi.org/10.1063/1.4746119.

[15]   E. Umbach, M. Sokolowski, R. Fink, and Epitaxy of Large Organic Adsorbates, Appl. Phys. A 63 (1996) 565–576.

[16]   C. Stadler, S. Hansen, I. Kröger, C. Kumpf, E. Umbach, Tuning intermolecular interaction in long-range-ordered submonolayer organic films, Nat. Phys. 5 (2009) 153–158. https://doi.org/10.1038/nphys1176.

[17]   Q. Wei, W. Liu, M. Leclerc, J. Yuan, H. Chen, Y. Zou, A-DA'D-A non-fullerene acceptors for high-performance organic solar cells, 63 (2020) 1352–1366.

[18]   J. Noh, M. Hara, Asymmetric Disulfide Self-Assembled Monolayers on, Thin Solid Films 16 (2000) 14–17.

[19]   N.T.N. Ha, A. Sharma, D. Slawig, S. Yochelis, Y. Paltiel, D.R.T. Zahn, G. Salvan, C. Tegenkamp, Charge-Ordered α-Helical Polypeptide Monolayers on Au(111), J. Phys. Chem. C 124 (2020) 5734–5739. https://doi.org/10.1021/acs.jpcc.0c00246.

[20]   A. Arjona-Esteban, J. Krumrain, A. Liess, M. Stolte, L. Huang, D. Schmidt, V. Stepanenko, M. Gsänger, D. Hertel, K. Meerholz, F. Würthner, Influence of Solid-State Packing of Dipolar Merocyanine Dyes on Transistor and Solar Cell Performances, J. Am. Chem. Soc. 137 (2015) 13524–13534. https://doi.org/10.1021/jacs.5b06722.

[21]   A. Zitzler-Kunkel, M.R. Lenze, N.M. Kronenberg, A.M. Krause, M. Stolte, K. Meerholz, F. Würthner, NIR-absorbing merocyanine dyes for BHJ solar cells, Chem. Mater. 26 (2014) 4856–4866. https://doi.org/10.1021/cm502302s.

[22]   A. Lv, M. Stolte, F. Würthner, Head-To-Tail Zig-Zag Packing of Dipolar Merocyanine Dyes Affords High-Performance Organic Thin-Film Transistors, Angew. Chemie - Int. Ed. 54 (2015) 10512–10515. https://doi.org/10.1002/anie.201504190.

[23]   A. V. Kulinich, A.A. Ishchenko, A.K. Chibisov, G. V. Zakharova, Effect of electronic asymmetry and the polymethine chain length on photoprocesses in merocyanine dyes, J. Photochem. Photobiol. A Chem. 274 (2014) 91–97. https://doi.org/10.1016/j.jphotochem.2013.09.016.

[24]   A. V. Kulinich, A.A. Ishchenko, U.M. Groth, Electronic structure and solvatochromism of merocyanines. NMR spectroscopic point of view, Spectrochim. Acta - Part A Mol. Biomol. Spectrosc. 68 (2007) 6–14. https://doi.org/10.1016/j.saa.2006.10.043.

[25]   S. Winkler, J. Frisch, P. Amsalem, S. Krause, M. Timpel, M. Stolte, F. Würthner, N. Koch, Impact of molecular dipole moments on fermi level pinning in thin films, J. Phys. Chem. C 118 (2014) 11731–11737. https://doi.org/10.1021/jp5021615.

[26]   S. Krause, M. Stolte, F. Würthner, N. Koch, Influence of merocyanine molecular dipole moments on the valence levels in thin films and the interface energy level alignment with au(111), J. Phys. Chem. C 117 (2013) 19031–19037. https://doi.org/10.1021/jp4060395.

[27]   K. Seki, H. Yanagi, Y. Kobayashi, T. Ohta, T. Tani, UV photoemission study of dye/AgBr interfaces in relation to spectral sensitization, Phys. Rev. B 49 (1994) 2760–2767. https://doi.org/10.1103/PhysRevB.49.2760.

[28]   T. Araki, E. Ito, K. Oichi, R. Mitsumoto, M. Sei, H. Oji, Y. Yamamoto, Y. Ouchi, K. Seki, Y. Takata, K. Edamatsu, T. Yokoyama, T. Ohta, Y. Kitajima, S. Watanabe, K. Yamashita, T. Tani, NEXAFS spectroscopic study of organic photographic dyes and their adsorbed states on AgCl and AgBr, J. Phys. Chem. B 101 (1997) 10378–10385. https://doi.org/10.1021/jp971934l.

[29]   S. Winkler, J. Frisch, P. Amsalem, S. Krause, M. Timpel, M. Stolte, F. Würthner, N. Koch, Impact of molecular dipole moments on fermi level pinning in thin films, J. Phys. Chem. C 118 (2014) 11731–11737. https://doi.org/10.1021/jp5021615.

[30] N.J. Hestand, F.C. Spano, Expanded Theory of H- and J-Molecular Aggregates: The Effects of Vibronic Coupling and Intermolecular Charge Transfer, Chem. Rev. 118 (2018) 7069–7163. https://doi.org/10.1021/acs.chemrev.7b00581.

[31] A.J. Kny, M. Reimer, N. Al-Shamery, R. Tomar, T. Bredow, S. Olthof, D. Hertel, K. Meerholz, M. Sokolowski, Chiral self-organized single 2D-layers of tetramers from a functional donor-acceptor molecule by the surface template effect, Nanoscale 15 (2023) 10319–10329. https://doi.org/10.1039/d3nr00767g.

[32] H. Bürckstümmer, E. V. Tulyakova, M. Deppisch, M.R. Lenze, N.M. Kronenberg, M. Gsänger, M. Stolte, K. Meerholz, F. Würthner, Efficient Solution-Processed Bulk Heterojunction Solar Cells by Antiparallel Supramolecular Arrangement of Dipolar Donor–Acceptor Dyes, Angew. Chemie 123 (2011) 11832–11836. https://doi.org/10.1002/ange.201105133.

[33] P.S. Bagus, V. Staemmler, C. Wöll, Exchangelike Effects for Closed-Shell Adsorbates: Interface Dipole and Work Function, Phys. Rev. Lett. 89 (2002) 1–4. https://doi.org/10.1103/PhysRevLett.89.096104.

[34] H. Vázquez, Y.J. Dappe, J. Ortega, F. Flores, Energy level alignment at metal/organic semiconductor interfaces: "Pillow" effect, induced density of interface states, and charge neutrality level, J. Chem. Phys. 126 (2007). https://doi.org/10.1063/1.2717165.

[35] G. Witte, S. Lukas, P.S. Bagus, C. Wöll, Vacuum level alignment at organic/metal junctions: "cushion" effect and the interface dipole, Appl. Phys. Lett. 87 (2005) 1–3. https://doi.org/10.1063/1.2151253.

[36] H. Yamane, Y. Yabuuchi, H. Fukagawa, S. Kera, K.K. Okudaira, N. Ueno, Does the molecular orientation induce an electric dipole in Cu-phthalocyanine thin films?, J. Appl. Phys. 99 (2006). https://doi.org/10.1063/1.2192978.

[37] T. Schultz, T. Lenz, N. Kotadiya, G. Heimel, G. Glasser, R. Berger, P.W.M. Blom, P. Amsalem, D.M. de Leeuw, N. Koch, Reliable Work Function Determination of Multicomponent Surfaces and Interfaces: The Role of Electrostatic Potentials in Ultraviolet Photoelectron Spectroscopy, Adv. Mater. Interfaces 4 (2017) 1–8. https://doi.org/10.1002/admi.201700324.

[38] H. Fukagawa, H. Yamane, T. Kataoka, S. Kera, M. Nakamura, K. Kudo, N. Ueno, Origin of the highest occupied band position in pentacene films from ultraviolet photoelectron spectroscopy: Hole stabilization versus band dispersion, Phys. Rev. B - Condens. Matter Mater. Phys. 73 (2006) 24–26. https://doi.org/10.1103/PhysRevB.73.245310.

[39] S. Yu, S. Ahmadi, C. Sun, K. Schulte, A. Pietzsch, F. Hennies, M. Zuleta, M. Göthelid, Crystallization-induced charge-transfer change in TiOPc thin films revealed by resonant photoemission spectroscopy, J. Phys. Chem. C 115 (2011) 14969–14977. https://doi.org/10.1021/jp1100363.

[40] F.J. Kahle, C. Saller, S. Olthof, C. Li, J. Lebert, S. Weiß, E.M. Herzig, S. Hüttner, K. Meerholz, P. Strohriegl, A. Köhler, Does Electron Delocalization Influence Charge Separation at Donor-Acceptor Interfaces in Organic Photovoltaic Cells?, J. Phys. Chem. C 122 (2018) 21792–21802. https://doi.org/10.1021/acs.jpcc.8b06429.

[41] S. Krause, M.B. Casu, A. Schöll, E. Umbach, Determination of transport levels of organic semiconductors by UPS and IPS, New J. Phys. 10 (2008). https://doi.org/10.1088/1367-2630/10/8/085001.

[42] M. Knupfer, H. Peisert, Electronic properties of interfaces between model organic semiconductors and metals, Phys. Status Solidi Appl. Res. 201 (2004) 1055–1074. https://doi.org/10.1002/pssa.200304332.

[43] N. Gildemeister, G. Ricci, L. Böhner, J.M. Neudörfl, D. Hertel, F. Würthner, F. Negri, K. Meerholz, D. Fazzi, Understanding the structural and charge transport property relationships for a variety of merocyanine single-crystals: A bottom up computational investigation, J. Mater. Chem. C 9 (2021) 10851–10864. https://doi.org/10.1039/d1tc01511g.

[44] A. Liess, A. Arjona-Esteban, A. Kudzus, J. Albert, A.M. Krause, A. Lv, M. Stolte, K. Meerholz, F. Würthner, Ultranarrow Bandwidth Organic Photodiodes by Exchange Narrowing in Merocyanine H- and J-Aggregate Excitonic Systems, Adv. Funct. Mater. 29 (2019) 1–9. https://doi.org/10.1002/adfm.201805058.

[45] J.H. Kim, T. Schembri, D. Bialas, M. Stolte, F. Würthner, Advanced Materials - 2021 - Kim - Slip-Stacked J-Aggregate Materials for Organic Solar Cells and Photodetectors.pdf, Adv. Mater. 34 (2022).

[46] R. Schäfer, L. Böhner, M. Schiek, D. Hertel, K. Meerholz, K. Lindfors, Strong Light-Matter Interaction of Molecular Aggregates with Two Excitonic Transitions, ACS Photonics 11 (2024) 111–120. https://doi.org/10.1021/acsphotonics.3c01042.

[47] Y. M., X-ray thin-film measurement techniques, Rigaku J. 167 (2010) 1–9. https://doi.org/10.1097/00010694-200208000-00001.

[48] K.A. Kacprzak, O. Lopez-Acevedo, H. Häkkinen, H. Grönbeck, Theoretical characterization of cyclic thiolated copper, silver, and gold clusters, J. Phys. Chem. C 114 (2010) 13571–13576. https://doi.org/10.1021/jp1045436.

[49] A.F. Carley, P.R. Davies, R. V. Jones, K.R. Harikumar, G.U. Kulkarni, M.W. Roberts, Structure of sulfur adlayers at Cu(110) surfaces: An STM and XPS study, Surf. Sci. 447 (2000) 39–50. https://doi.org/10.1016/S0039-6028(99)01191-7.

# Chemical interaction and molecular growth of a highly dipolar merocyanine molecule on metal surfaces: A photoelectron spectroscopy study

Baris Öcal,[1] Philipp Weitkamp,[1] Klaus Meerholz,[1] and Selina Olthof [1,*]

[1] Department Chemistry, Institute of Physical Chemistry, University of Cologne, Cologne 50939, Germany

Email: solthof@uni-koeln.de

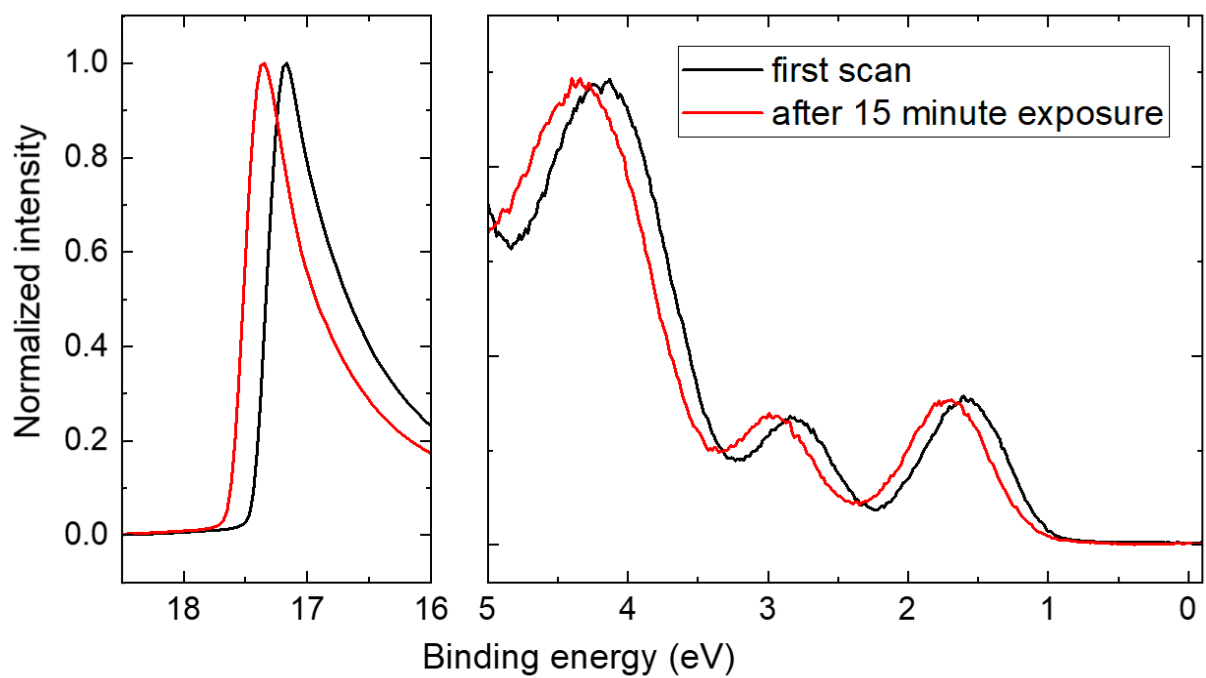

Figure S1: Degradation of the molecule HB238 under extended UV irradiation. As indicated in the experimental section, care has to be taken not to affect the organic layer during measurement. As shown here for 100 Å HB238 on Au(100), shifts for both high energy cutoff and the HOMO region under long UV light exposure can be seen. Typically, this shift only occurred for thickness above 5 nm.

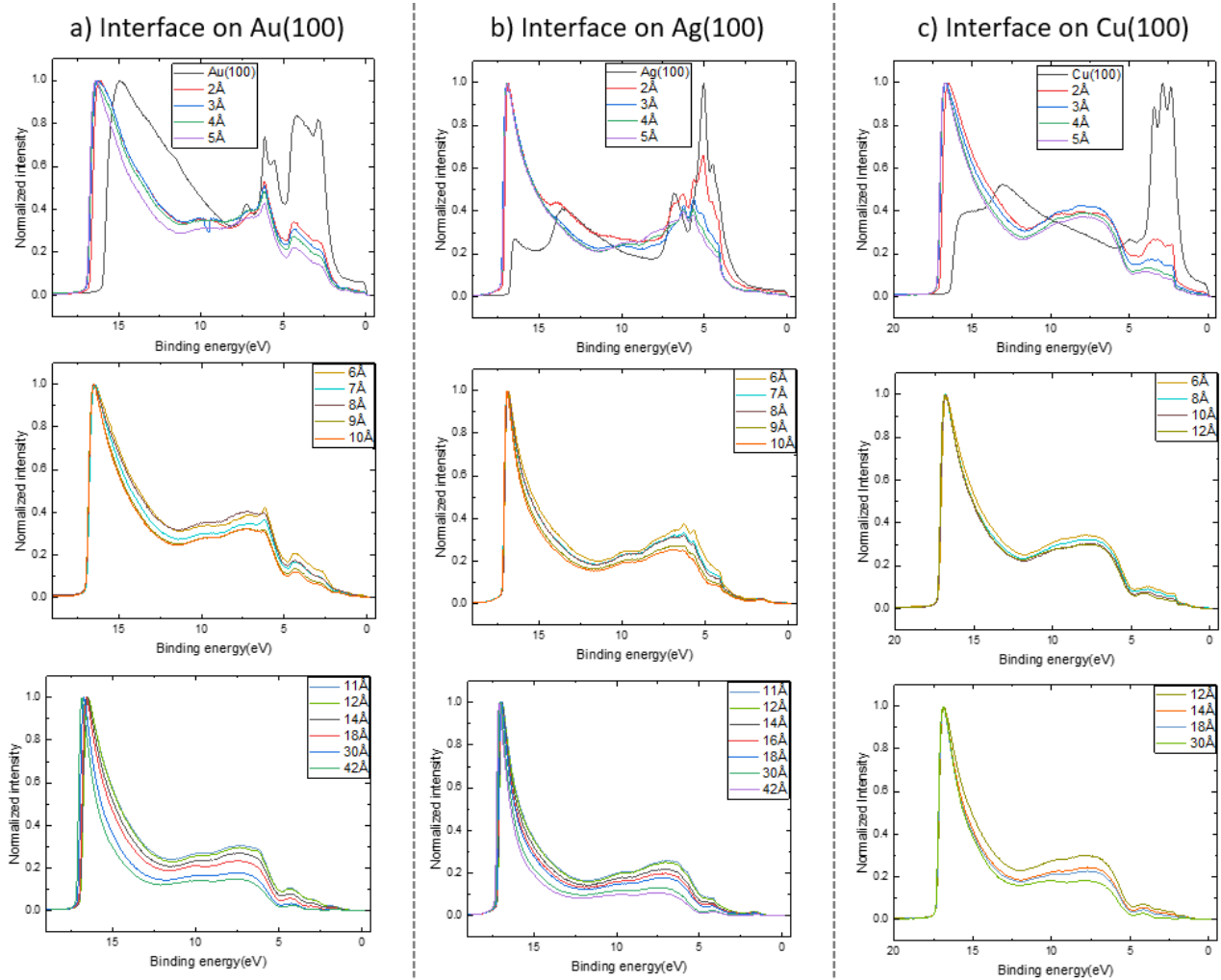

Figure S2: UPS scans of the different thicknesses of HB238 on a) Au(100), b) Ag(100) and c) Cu(100). For each interface analysis, the data set has been separated into three graphs to increase readability.

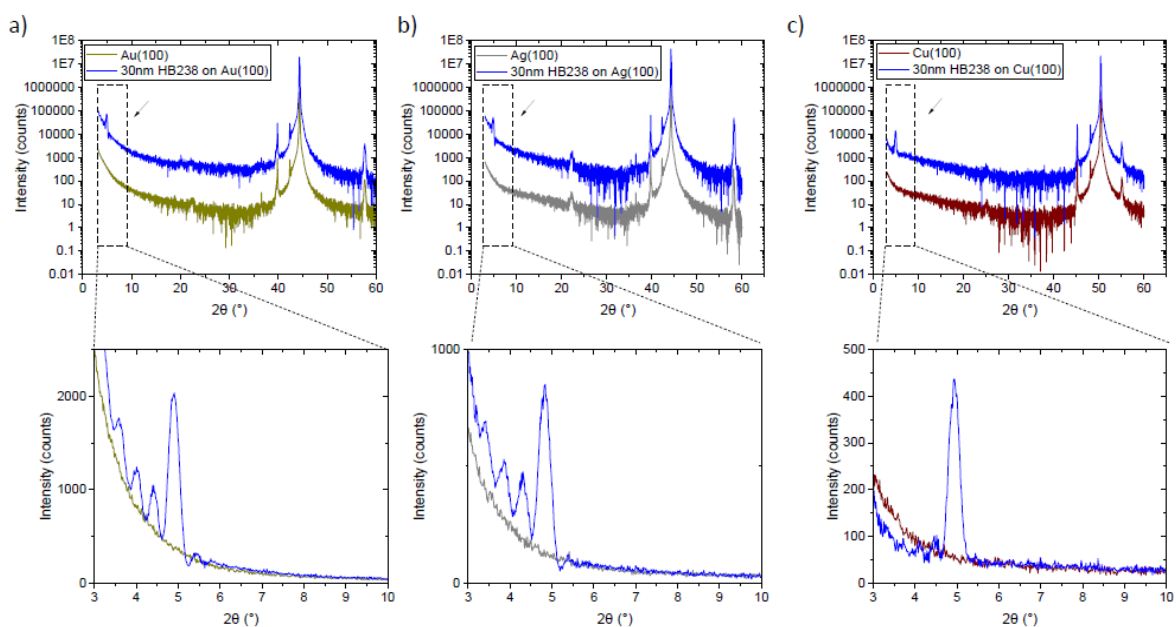

Figure S3: XRD patterns of the pure clean single crystal as well as after coverage with 30 nm HB238 on a) Au(100), b) Ag(100) and c) Cu(100). The **top row** shows the 2θ region between 3 and 60° in a semi-logarithmic scale, in order to visualize, both, the weak reflexes from the thin organic layer at low 2θ values, as well as the substrate related peaks at higher angle. The two curves in each panel have been vertically separated to be able to discern the two spectra. In the **bottom row**, the same spectra are shown in the 2Θ range between 3 and 10°. The Kiessing fringe intensities suggest that the film roughness is Cu(100) > Ag(100) > Au(100). Due to the different film roughness, the intensity of the molecules reflex at 2θ ~ 5° cannot be used to deduce the degree of crystallinity in these films. The correlated crystal structure from which the tip-on orientation of the molecules can be deduced, has been published by Gildemeister et al. [1]

Table S1: S2p$_{3/2}$ core level peak positions of HB238 for the different layer thicknesses on the three substrates Au(100), Ag(100) and Cu(100)

| | Au(100) | | Ag(100) | | Cu(100) | | |
|---|---|---|---|---|---|---|---|
| Layer thickness | S_donor | S_acceptor | S_donor | S_acceptor | S_donor | S_acceptor | S_metal bond |
| 2 | 164.32 | 164 | 164.14 | 164.11 | 163.7 | 163.67 | 161.67 |
| 3 | 164.17 | 163.86 | 164.24 | 163.86 | 163.68 | 163.67 | 161.67 |
| 7 | 164.47 | 163.83 | 164.17 | 163.78 | 164.71 | 163.85 | 161.67 |
| 11 | 164.51 | 163.85 | 164.36 | 163.76 | 164.65 | 163.92 | 161.67 |
| 42 | 164.64 | 164.09 | 164.51 | 163.96 | 164.59 | 164 | 161.67 |

Table S2: core level peak positions of HB238 for the different layer thicknesses on the three substrates Au(100), Ag(100) and Cu(100)

| | Au(100) | | | Ag(100) | | | Cu(100) | | |
|---|---|---|---|---|---|---|---|---|---|
| Layer thickness | N(1) | N(2) | N(3) | N(1) | N(2) | N(3) | N(1) | N(2) | N(3) |
| 2 | 400.23 | 399.03 | 397.97 | 400.2 | 399.14 | 398.19 | 399.46 | 398.8 | 397.96 |
| 3 | 400.19 | 399.03 | 397.92 | 399.74 | 399.02 | 398.12 | 399.46 | 398.7 | 398 |
| 7 | 400.19 | 399.11 | 398.14 | 399.59 | 398.63 | 398.12 | 399.66 | 398.83 | 398.1 |
| 11 | 400.08 | 399.14 | 398.26 | 399.74 | 398.76 | 398.01 | 399.89 | 398.97 | 398.14 |
| 42 | 400.11 | 398.99 | 398.33 | 400 | 398.79 | 398.11 | 400.1 | 398.96 | 398.21 |

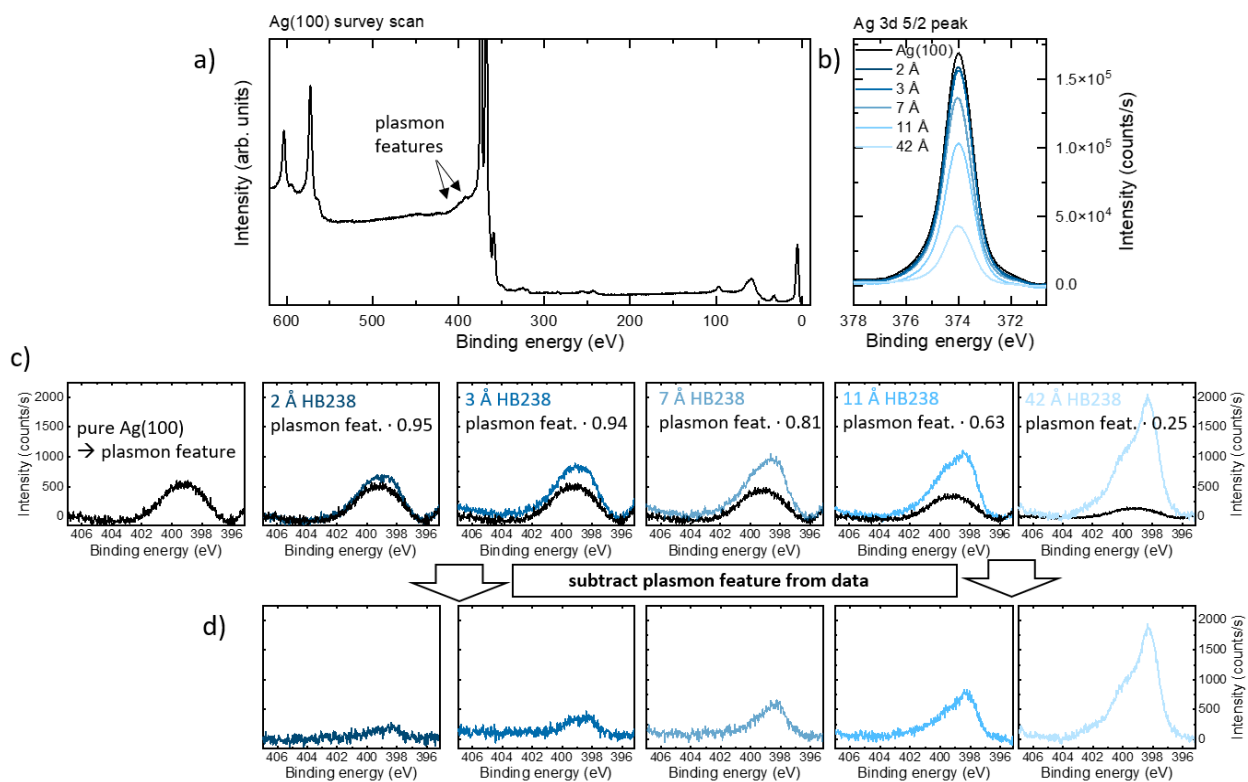

Figure S4: Background subtraction procedure for the N1s region of the sample series prepared on Ag(100). a) shows the survey spectrum of clean Ag(100) in which the plasmon features related to the Ag 3d core level signal are indicated. These partially coincide with the N1s region and have to be subtracted to be able to fit the weak nitrogen signal of the HB238 molecule. This Ag plasmon-related background signal is shown in the first panel of c) in black color. The additional panels in c) show the same binding energy region upon deposition of the molecule HB238, as indicated. To obtain the pure N1s signals, the remaining Ag plasmon signal needs to be subtracted. To estimate the intensity of this plasmon peak, the Ag $3d_{5/2}$ peak has been measured (shown in b) and the plasmon signal is scaled to the same intensity as this core level peak; the scaling factor is given in each of the panels of c). The resulting pure N1s spectra after plasmon peak subtraction are given in sub-Figure d) and are the ones used in Figure 4b1) of the main article to fit the nitrogen signal.

## Additional References

[1] N. Gildemeister, G. Ricci, L. Böhner, J.M. Neudörfl, D. Hertel, F. Würthner, F. Negri, K. Meerholz, D. Fazzi, *Understanding the structural and charge transport property relationships for a variety of merocyanine single-crystals: A bottom up computational investigation*, J. Mater. Chem. C, 9 (2021) 10851. DOI: 10.1039/d1tc01511g.